\documentclass[twoside,12pt]{article}
\usepackage{CJK}
\usepackage{indentfirst}
\usepackage{bm}
\usepackage{epsfig}
\usepackage{color}

\usepackage{graphicx}
\usepackage{physics}
\usepackage{caption}
\usepackage{subcaption}

\begin{document}

\begin{CJK*}{GBK}{song}

\thispagestyle{empty} \vspace*{0.8cm}\hbox
to\textwidth{\vbox{\hfill\huge\sf Commun. Theor. Phys.\hfill}}
\par\noindent\rule[3mm]{\textwidth}{0.2pt}\hspace*{-\textwidth}\noindent
\rule[2.5mm]{\textwidth}{0.2pt}


\begin{center}
\LARGE\bf {$^4$He monolayer on Graphene:\\ a Quantum Monte Carlo study}
\end{center}

\footnotetext{\hspace*{-.45cm}\footnotesize $^\dag$Corresponding author, E-mail: m.boninsegni@ualberta.ca }

\begin{center}
\rm S. Yu \ and  \ M. Boninsegni$^\dagger$
\end{center}

\begin{center}
\begin{footnotesize} \sl
Department of Physics, University of Alberta, Edmonton, Alberta, Canada T6G 2H5
\end{footnotesize}
\end{center}

\begin{center}
\footnotesize (Received 2 March 2024; revised manuscript received 16 May 2024)

\end{center}

\vspace*{2mm}

\begin{center}
\begin{minipage}{15.5cm}
\parindent 20pt\footnotesize
We revisit the problem of adsorption of a single $^4$He layer on graphene, focusing on the commensurate (C$_{1/3}$) crystalline phase, specifically 
on whether it may possess a nonzero superfluid response, and on the existence of superfluid phases, either (metastable) liquid or vacancy-doped crystalline. We make use of canonical Quantum Monte Carlo simulations at zero and finite temperature, based on a realistic microscopic model of the system. Our results confirm the absence of any superfluid response in the commensurate crystal, and that no thermodynamically stable uniform phase exists at lower coverage. No evidence of a possibly long-lived, metastable superfluid phase at $C_{1/3}$ coverage is found. Altogether, the results of ground-state projection methods and finite-temperature simulations are entirely consistent. 
\end{minipage}
\end{center}
\bigskip\bigskip
\begin{center}
\begin{minipage}{15.5cm}
\begin{minipage}[t]{2.3cm}{\bf Keywords:}\end{minipage}
\begin{minipage}[t]{13.1cm}
Quantum fluids and solids; Quantum Monte Carlo; Two-dimensional systems
\end{minipage}\par\vglue8pt

\end{minipage}
\end{center}
\newpage
\section{Introduction}
\label{introd}
The physics of thin helium films varies greatly, depending on the substrate upon which they are adsorbed \cite{Taborek2020}. The equilibrium phase of $^4$He in two dimensions (2D) is a superfluid liquid, and indeed on the weakest substrate for which a stable 2D film forms, namely lithium, a stable superfluid monolayer is predicted \cite{Boninsegni1999} and observed \cite{VanCleve2008}, with continuous growth of film as a function of the chemical potential \cite{Boninsegni2004}. On the other hand, on stronger substrates such as graphite, the phase diagram features a dazzling variety of phases, including crystalline ones, either commensurate or incommensurate with the underlying substrate. 

An interesting theoretical question is whether a phase simultaneously displaying diagonal (i.e., crystalline) and off-diagonal (i.e., superfluid) long-range order may occur at sufficiently low temperature, in some well-defined range of $^4$He coverage. 
Some experimental claims have been made of concurrent superfluidity and crystalline order in the second layer of $^4$He adsorbed on graphite, specifically at, or in the vicinity, of a solid phase registered with the underlying first (solid) layer \cite{Crowell1993,Nakamura2016,Nyeki2017,Choi2021}. Though the denomination {\em supersolid} is, strictly speaking, not applicable to a system of this type \cite{Boninsegni2012,Boninsegni2012e}, the observation of such a phase would nonetheless be of great significance, as it would help elucidate the interplay between these two seemingly antithetic types of order. It seems fair to state, however, that at this time all the evidence is indirect and/or inconclusive; indeed, the interpretation of the findings has been questioned and is not supported by reliable microscopic theoretical calculations \cite{Corboz2008,Kwon2016,Moroni2019,Boninsegni2020,Boninsegni2023}.

Graphene, which consists of a sheet of graphite, is also regarded as a potentially interesting adsorber. A single layer of carbon atoms is about 10\% less attractive than graphite; the ensuing enhancement of the quantum excursions of helium atoms away from the basal plane acts to soften the repulsive interaction among them at short distances. All of this may ultimately increase the mobility of $^4$He, possibly stabilizing novel phases displaying a more quantal character (e.g., superfluid phases with an unusual degree of atomic localization).  Early microscopic calculations for helium on graphite showed that a mere 10\% reduction in the strength of the attractive part of the interaction potential results in no significant change in the overall physical behavior of an adsorbed helium film \cite{Corboz2008}, a conclusion confirmed by subsequent calculations for $^4$He on graphene. In particular, there is general agreement on the existence of the same commensurate crystalline phase of a $^4$He monolayer (known as $C_{1/3}$) which forms on graphite. 
Indeed, the basic features of the phase diagram of an adsorbed $^4$He monolayer on graphene can be understood on the basis of a highly simplified description of the system, making use of a lattice Hamiltonian of hard core bosons \cite{Yu2021}.

There is, however, some controversy concerning a possible finite superfluid response of such a crystalline phase, at sufficiently low temperature, as well as on the existence of superfluid phases at lower coverage, either liquid or crystalline (doped with vacancies).
\\
Finite-temperature calculations have yielded no evidence of a nonzero superfluid response of the $C_{1/3}$ phase \cite{Kwon2012,Happacher2013}, while the claim of a small superfluid signal in the ground state has been made \cite{soliti2011}.
Furthermore, some studies \cite{Happacher2013,soliti2009} have given evidence of a possible low-lying (metastable) superfluid phase at sufficiently low temperature, one that could conceivably be long-lived, and therefore observed experimentally.

In this work, we revisit the issue of superfluidity at low temperature of a monolayer of $^4$He absorbed on a single graphene sheet, at and below commensurate coverage, by means of Quantum Monte Carlo (QMC) simulations at zero and finite temperature. We make use of a microscopic model of the system that is consistent with most previous studies, explicitly accounting for the corrugation of the graphene substrate. We monitor the superfluid response and the emergence of diagonal long-range order at low temperature. 

Altogether, our results confirm the absence of superfluidity in this system at commensurate coverage, all the way to temperature $T=0$. No inconsistency is observed between finite-temperature and ground-state simulations, i.e., even the latter yield no evidence of any finite superfluid response. In particular, our ground state estimates result from projecting the lowest-energy component from an initial trial wave function that does not break translational invariance, i.e., with no built-in information about the presence of a corrugated substrate. Yet, crystalline long-range order quickly builds in, even for relatively short projection imaginary-time intervals, disproving the contention put forth in Ref. \cite{soliti2009} about the existence of a superfluid state energetically competitive with the crystalline ground state, based on a similar projection procedure. 
\\ \indent
We find no evidence of a thermodynamically stable, uniform superfluid phase for coverage below commensuration; indeed, our results are in disagreement with the contention made in Ref. \cite{Kwon2012} that structural and  superfluid properties of the system below commensuration are greatly affected by the choice of potential describing the interaction of a helium and a carbon atom, specifically an isotropic form of the potential leading to a superfluid uniform phase below commensurate coverage. On the contrary, we find the physical picture to be the same as that predicted with an anisotropic potential, with commensurate crystalline order persisting in the presence of vacancies, and with the formation of solid clusters at low coverage, analogously to what has been predicted to occur on graphite. 
Finally, we see no evidence of any metastable superfluid phase at commensurate coverage.

This manuscript is organized as follows: we introduce the microscopic model of the physical system for which we have carried out our calculations and describe the computational methodology adopted in this work in section 2, 
present our results in detail in section 3, and outline our conclusion in section 4.

\section{Model and methodology}
\label{modham}
We consider an assembly of $N$ $^4$He atoms, regarded as point-like identical particles of mass $m$ and spin $S=0$,  moving in three physical dimensions.
The system is enclosed in a cell of volume  $\Omega=L_x\times L_y\times L_z$, with $L_x=34.08$ \AA, $L_y = 29.514$ \AA, and $L_z=40$ \AA. Periodic boundary conditions are used in the three directions (but the boundary condition along the $z$ direction is immaterial, as helium atoms remain confined in the vicinity of the substrate). The cell sizes are chosen to accommodate an integer number of graphene unit cells in both directions, without introducing frustration. The graphene sheet is modeled as an ideal, 2D (honeycomb) lattice; our simulation cell accommodates $N_c=384$ carbon atoms at fixed positions on the $z=0$ plane, with a carbon-carbon bond length $a = 1.42$ \AA. Regarding the carbon atoms as pinned at classical lattice sites is an assumption that has been made in all other comparable studies, and one that seems justified by their relatively large mass, compared to that of the helium atoms. The $^4$He coverage (i.e., 2D density) is $\theta=N/A$, where $A=L_x\times L_y$.

The quantum-mechanical many-body Hamiltonian reads as follows:
\begin{equation}\label{u}
\hat H =   \sum_{i}\biggl [-\lambda \nabla^2_{i}+U({\bf r}_i)\biggr ] +\sum_{i<j}v(r_{ij})
\end{equation}
where the first (second) sum runs over all particles (pairs of particles), $\lambda\equiv\hbar^2/2m=6.05964$ K\AA$^{2}$, $r_{ij}\equiv |{\bf r}_i-{\bf r}_j|$, $v(r)$ denotes the pairwise interaction between two $^4$He atoms and
\begin{equation} U({\bf r})=\sum_{\alpha=1}^{N_c} u(|{\bf r}-{\bf R}_\alpha|),\end{equation}
the sum running over all carbon atoms and $u$ being a potential describing the interaction between a helium and a carbon atom. Both $v$ and $u$ are assumed to depend only on the relative distance between two particles; this is a standard assumption for the helium pair potential $v$, justified by the essentially spherical shape of a helium atom in its ground state. For the helium-carbon part, there is still uncertainty on whether a spherically symmetric potential $u$ is adequate. \\ \indent
Because the individual carbon atoms are explicitly included in Eq. \ref{u}, the model is expected to capture the most important contribution of the corrugation of the graphene substrate, which is necessary to stabilize a commensurate crystalline layer \cite{Whitlock1998}. However, Kwon and Ceperley \cite{Kwon2012} have contended that the physics of the system below the commensurate $C_{1/3}$ coverage $\theta_0=0.0636$  \AA$^{-2}$ is significantly affected by the choice of $u$; in particular, a superfluid phase (either liquid or crystalline, doped with vacancies) exists if an isotropic form of the potential $u$ is used, while, on using an anisotropic form, no thermodynamically stable phase is found for coverage $\theta < \theta_0$. This assertion is at variance with the subsequent findings of Happacher {\em et al.} \cite{Happacher2013}, who carried out grand canonical simulations using an isotropic potential $u$ and found no evidence of any thermodynamically stable phase of coverage less than $\theta_0$.
\\ \indent
The choice made here for $u$ and $v$ is motivated both to ensure consistency with (most of) the existing calculations, as well as to clarify some of the above, still unresolved aspects of the physics of the system for the simplest choice of helium-carbon interaction, i.e., that based on an isotropic form for $u$.
We adopt therefore the same version of the Aziz potential utilized in Refs. \cite{soliti2011,soliti2009}, though it should be noted that the differences between the various versions of the Aziz potential are quantitatively small and that the helium-graphene interaction dominates the energetics in this system. The carbon-helium pair-wise interaction $u$ is a Lennard-Jones potential with parameters $\epsilon=16.25$ K and $\sigma=2.74$ \AA\ taken from Ref. \cite{Carlos1980}, again for consistency with previous calculations. In our simulation, the potentials are cut off and smoothly shifted to zero at a distance $r_c=14.75$ \AA; we estimate the energetic contribution arising from interactions between particles located at greater distances to amount to 0.36 K per $^4$He atom.
\\ \indent
The low-temperature phase diagram of the thermodynamic system described by Eq.  (\ref{u}) as a function of coverage and temperature has been studied in this work by computer simulations at zero and finite temperature. In the majority of our calculations the coverage $\theta=\theta_0$, i.e., it is $N=64$. For this coverage, the $C_{1/3}$ crystalline monolayer $^4$He film commensurate with the underlying graphene substrate is expected to form.
We also carried out some simulations at lower coverage, to investigate the possible occurrence of a thermodynamically stable superfluid phase, either liquid or a vacancy-rich commensurate crystal. 
\\ \indent
Finite-temperature simulations have been carried out in the temperature range $0.25 \le T \le 4$ K. They are based on the continuous-space Worm Algorithm \cite{Boninsegni2006,Boninsegni2006b}.  Since this technique is well-established, and extensively described in the literature, we shall not review it here; we used the canonical variant of the algorithm, in which the number of particles $N$ is fixed \cite{Mezzacapo2006,Mezzacapo2007}. The most important aspects of this methodology, in the context of the study carried out here are that it allows one to obtain  unbiased estimates of energetic and structural properties of the system, essentially with no approximation, as well as of the superfluid fraction, based on the winding number estimator \cite{Pollock1987}.
\\ \indent
Ground state simulations, on the other hand, were performed using the related Path Integral Ground State (PIGS) method \cite{Sarsa2000}, which projects the true ground state out of an initial trial wave function.
Details of both simulations are standard; in both cases, we made use of a fourth-order approximation for the short imaginary time ($\tau$)  propagator (see, for instance, Refs. \cite{Cuervo2005,Boninsegni2005}), and all of the results presented here are extrapolated to the $\tau\to 0$ limit. We generally found numerical estimates for the physical properties of interest here obtained with a value of the time step $\tau\sim 1.5\times 10^{-3}$ K$^{-1}$ to be indistinguishable from the extrapolated ones, within the statistical uncertainties of the calculation (the only exception is the energy in ground state calculations, for which a smaller value of the time step was needed. We discuss this aspect in greater depth below).

\section{Results}\label{res}
As a preliminary comment, which applies to both finite-temperature and ground-state results presented here, it needs to be mentioned that in all cases they are independent of the initial atomic
configuration utilized in the simulations.
\subsection{Finite-temperature results}
\begin{figure}[t]
\center
\includegraphics*[width=0.7\linewidth]{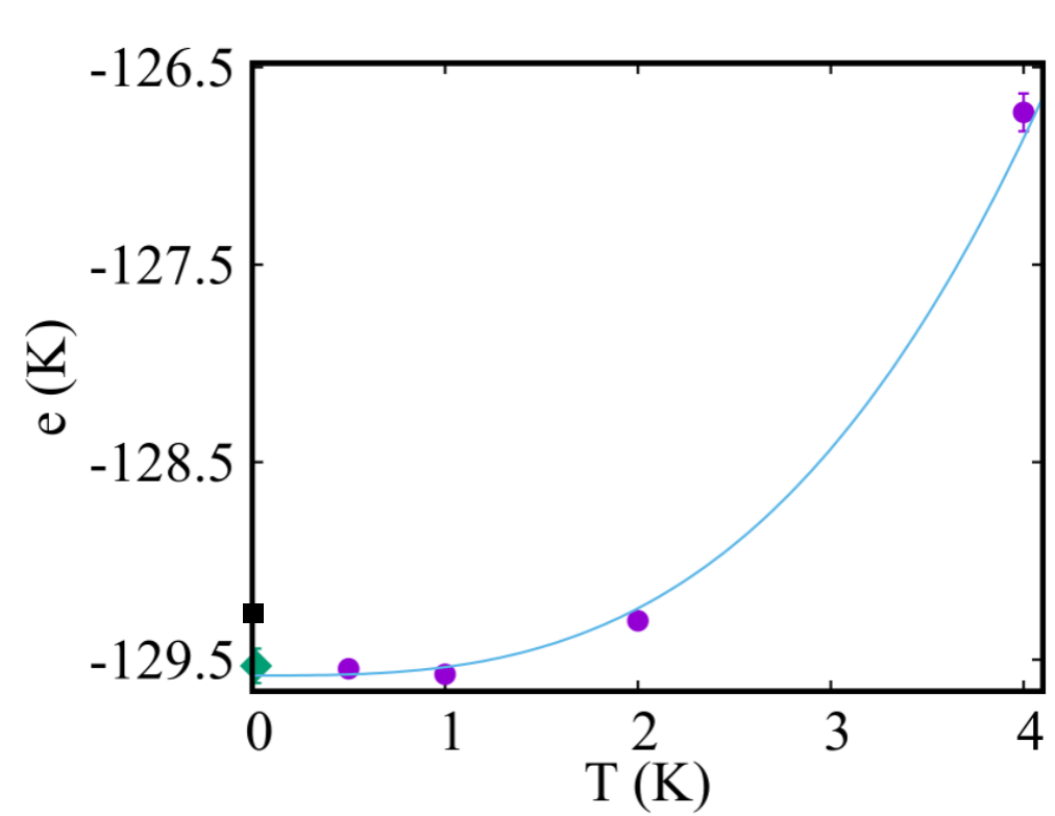}
\caption{ Energy per $^4$He atom (in K) computed by simulation as a function of the temperature at a coverage $\theta=0.0636$ \AA$^{-2}$. Statistical errors are smaller than symbol sizes. The solid line represents a fit to the data obtained with the expression $e(T)=e_0+\alpha T^3$. The estimate shown with a diamond at $T=0$ is that obtained with PIGS (see section 3.2) for a projection time $\Lambda=$\ 1 K$^{-1}$, while the black box refers to the estimate of Ref. \cite{soliti2009}. When not shown, statistical errors are smaller than symbol sizes.}
\label{energy}
\end{figure}

We begin with a discussion of the low-temperature energetics of the system. Fig. \ref{energy}  shows the computed energy per $^4$He atom $e(T)$ as a function of temperature, at the commensurate coverage $\theta_0$. As shown in Fig. \ref{energy}, the estimates are unchanged below $T=1$ K, i.e., they can be regarded as ground-state results.
The data can be fitted with the expression $e(T)=e_0+\alpha T^3$, yielding $e_0=-129.550 (25)$ K. The first observation is that energy estimates at finite temperature (for $T\le 1$ K), for the particular coverage considered here are 
significantly {\em lower} (by about 0.25 K) than the {\em ground state} estimate quoted in  Ref. \cite{soliti2009}, based on a calculation making use of the same potentials utilized in this work \cite{potential}. \\ \indent
It is worth noting that the difference between our estimate and that of Ref. \cite{soliti2009} is three times greater than that between the two results provided in Ref. \cite{soliti2009}, projecting the lowest-energy state out of two initial trial wave functions, one possessing crystalline long-range order and the other not breaking translational invariance. It was argued therein that the relatively small energy difference arising from the calculation carried out with the two different trial wave functions pointed to the existence of a low-lying superfluid phase, energetically competitive with the insulating, crystalline ground state.
The results obtained in this work suggest instead that the (Diffusion Monte Carlo, DMC) projection utilized in Ref. \cite{soliti2009} actually {\em failed} to reach convergence to the ground state with {\em either} trial wave function \cite{comment}. The same observation was made in the past \cite{Boninsegni2013}, pointing out how path integral approaches, which are not affected by population control bias \cite{Baroni1999,Boninsegni2001,Boninsegni2012b} are typically a more reliable option than DMC, when it comes to investigating the ground state of Bose systems. 
\begin{figure}[t]
\center
\includegraphics*[width=1. \linewidth]{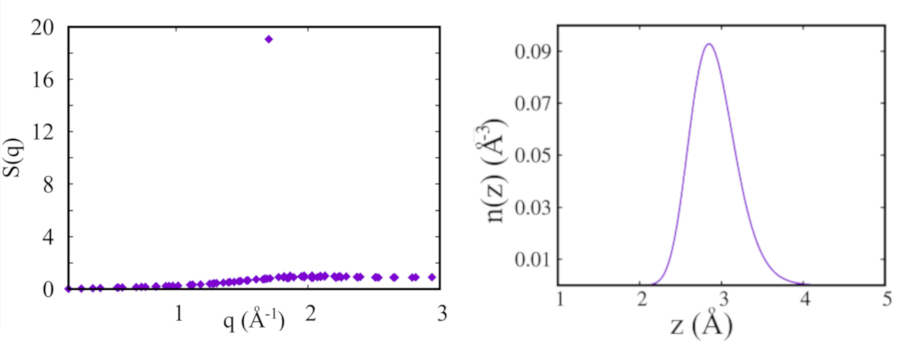}
\caption{{\em Left:} Static structure factor $S(q)$ for an adsorbed film of $^4$He on graphene, computed by simulation at temperature $T=0.5$ K, for wave vectors all lying in the graphene plane. {\em Right}: Average $^4$He density as a function of the distance from the graphene plane, computed at the same temperature.}
\label{static}
\end{figure}

The occurrence of crystalline long-range order at low temperature at coverage $\theta_0$ can be quantitatively established by the calculation of the static structure factor $S(q)$, shown in Fig. \ref{static} (left). The occurrence of a sharp (Bragg) peak at $q=1.7$ \AA$^{-1}$ is what one expects for the commensurate ($\sqrt 3 \times \sqrt 3$) commensurate crystal. The right part of Fig. \ref{static} shows instead the average $^4$He density profile as a function of the distance from the graphene plane. The results shown for both quantities are for a temperature $T=0.5$ K, but they are found to be essentially temperature-independent below $T=2$ K. 

No evidence of superfluidity is seen at the lowest temperature considered here, namely 0.5 K, with exceedingly infrequent atomic exchanges, never extending beyond a four-atom ring permutation. Obviously, however, the onset of superfluidity should be expected at a considerably lower temperature than 0.5 K. Indeed, the estimate of the superfluid fraction proposed in Ref. \cite{soliti2009} would suggest a superfluid transition temperature of the order of 5 mK, based on the well-known ``universal jump'' condition \cite{Nelson1977}. We come back to a critical re-examination of this, and other claims made in Ref. \cite{soliti2009} below, when discussing our ground state results. For the moment, we simply note the paucity of atomic exchanges, which instead occur quite frequently at this temperature in 2D, as well as that a previous study of this system, based on the same computational methodology utilized here yielded no evidence of superfluidity of the $C_{1/3}$ phase down to temperatures as low as 10 mK \cite{Happacher2013}. 
\subsubsection{Low coverage results}
In Ref. \cite{Kwon2012} the claim was made that, upon using an isotropic helium-carbon potential in model (\ref{u}), one could observe a (metastable) superfluid phase at coverage below $\theta_0$, typically by introducing a few vacancies, or by reducing the $^4$He coverage to approximately 0.058 \AA$^{-2}$. 
Our study yielded no evidence of any low-coverage superfluid liquid phase. Specifically, our simulations at coverage $\theta\le \theta_0$ unambiguously show crystalline order, consistently with the coexistence of a crystal (of coverage $\theta_0$) and vapor, which is in agreement with the findings of Ref. \cite{Happacher2013}. The claim of a liquidlike phase of Ref. \cite{Kwon2012}  is based on the observation of a sudden drop of the peak of the static structure factor, on lowering the coverage from $\theta_0$ to approximately 0.9 $\theta_0$, and from the visual inspection of the configuration generated in the course of a Monte Carlo simulation, suggesting loss of local crystalline order at the lower coverage. On the other hand, if an anisotropic pair potential is utilized the static structure factor does not drop as significantly, and local crystalline order is retained. The results of our simulations show that, while, on the one hand, the anisotropic potential can plausibly be more effective at ``pinning'' helium atoms in place than the isotropic one, the main physical behavior is actually unchanged in the case of an isotropic potential.
\begin{figure}[t]
\center
\includegraphics*[width=1. \linewidth]{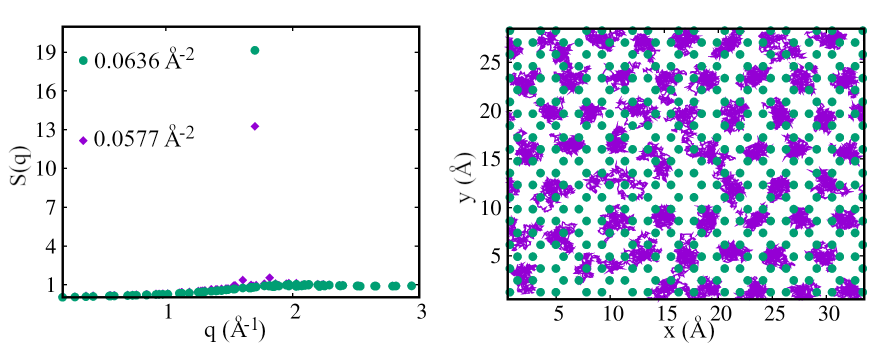}
\caption{{\em Left:} Static structure factor $S(q)$ for an adsorbed film of $^4$He on graphene, computed by simulation at temperature $T=1$ K, for wave vectors all lying in the graphene plane, at coverage $\theta_0=0.0636$ \AA$^{-2}$ (circles) and $\theta=0.0577$ \AA$^{-2}$ (diamonds). Statistical errors are smaller than symbol sizes. {\em Right}: Snapshot of many-atom configuration (particle world lines) corresponding to the lower coverage. Solid circles represent the carbon atoms.}
\label{lowtheta}
\end{figure}

Fig. \ref{lowtheta} shows a comparison of the static structure factor computed at $T=1$ K (the main results and observations do not change at lower temperatures, down to the lowest considered here, namely $T=0.25$ K) for the case of coverage $\theta_0$ (circles) and $\theta=0.0577$ \AA$^{-2}$ (diamonds). The peak of $S(q)$ is lower at the lower coverage (though the difference is nowhere near as large as that reported in Ref. \cite{Kwon2012}), but this is insufficient to conclude that no crystalline order is present at the lower coverage. For example, the calculation of Ref. \cite{Kwon2012} at coverage $\theta_0$ yields a $\sim 50\%$ higher peak of $S(q)$ if an anisotropic He-C pair potential is used, but in both cases the crystalline character of the monolayer is not in question.  Furthermore, although indeed the main peak is depressed by approximately a third (left part of Fig. \ref{lowtheta}), nevertheless it remains relatively sharp, suggesting that crystalline order is largely retained, as confirmed by snapshots of instantaneous many-particle world line configurations (right part of Fig. \ref{lowtheta}).
\\ \indent
It is contended in Ref. \cite{Kwon2012} that, at coverage less than $\theta_0$, permutations of identical helium atoms begin to occur,  leading at sufficiently low temperature to a finite superfluid response, which for sufficiently low coverage is interpreted as the stabilization of a superfluid liquid-like state. In our simulations we also observe such permutations, in fact even at temperatures as high as 1 K (as shown in Fig. \ref{lowtheta}). They typically involve atoms that lack some nearest neighbors, and in a small simulation cell cycles involving several particles occasionally appear,  connecting two opposite sides of the cell and resulting in a significant superfluid signal, which, for relatively small system sizes (e.g., $\sim$ 30 atoms) may appear as a bulk superfluid response. This is a finite-size effect, however; in the thermodynamic limit coexistence of a crystal with a low-density vapor is expected. In all of our simulations, the estimate of the superfluid fraction obtained as an average over sufficiently long runs is zero within statistical errors (i.e., the average value is typically much smaller than the statistical error). This is true both at commensurate and at lower coverage.\\ \indent
Altogether, therefore, although the anisotropic He-C potential has a stronger pinning effect on the helium atoms, nonetheless the basic physics of the system below commensurate coverage is the same, regardless of whether an isotropic or an anisotropic potential is used. The physical picture that emerges from our simulations below commensurate coverage is the formation of solid $^4$He clusters, analogously to what is observed on graphite \cite{Pierce1999}.

\subsubsection{Search for metastable superfluid state}
In order to explore the possible existence of a metastable, long-lived liquid-like superfluid phase of coverage $\theta_0$ we have carried out the same ``computer experiment'' of Refs. \cite{Boninsegni2006c,Boninsegni2018}. Specifically, we initially prepared the system in a superfluid phase by simulating a $^4$He monolayer adsorbed on a Li substrate (using the same simulation cell and number of atoms). The Li substrate is modeled as flat (i.e., smooth, featureless); on it, at a temperature $T=0.5$ K, a $^4$He monolayer at coverage $\theta_0$ forms a nearly 2D superfluid, with a superfluid fraction close to 100\% \cite{Boninsegni1999}. Upon establishing that thermal equilibrium was reached, the lithium was replaced with the graphene substrate, i.e., we reverted to the Hamiltonian (\ref{u}), to assess the resilience of the superfluid phase. One would expect that if a low-lying metastable superfluid phase existed, the simulation algorithm may remain ``stuck'' in it, finding the true equilibrium (crystalline) phase requiring the disentanglement \cite{Boninsegni2012d} of the many-particle paths present in the low-temperature superfluid phase stabilized on the Li substrate.

\begin{figure}[t]
\center
\includegraphics*[width=1. \linewidth]{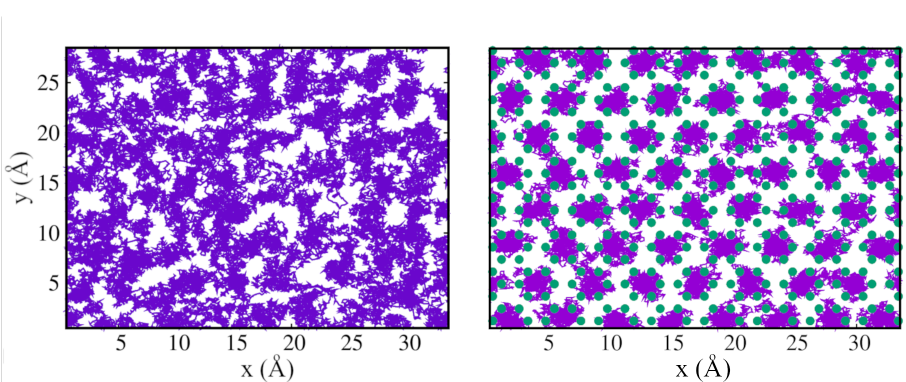}
\caption{Snapshots of many-particle configurations (world lines) corresponding to simulations of $^4$He on a flat Li substrate (left) and on the graphite substrate (right, small circles represent the C atoms). Both simulations are at $T=0.5$ K. The one on Li is such that the system is entirely superfluid, within the statistical errors of the calculation. The simulation on graphene is restarted from a configuration for the system on the left, after thermodynamic equilibrium is reached, simply with the replacement of the substrate.}
\label{compare}
\end{figure}

But what is observed, instead, is that as long as the all-important ``swap'' moves \cite{Boninsegni2006b} are attempted sufficiently often, the replacement of the Li substrate with the graphene causes the existing long atomic exchange cycles to disappear rather quickly, with a corresponding decrease of the initial superfluid signal. As most long permutation cycles have disappeared, the system {\em spontaneously} begins to develop solid order. All of this is illustrated in Fig. \ref{compare}, which displays instantaneous many-particle configurations (world lines) for the simulations conducted on a Li substrate (left part of Fig. \ref{compare}) and on the graphene substrate (right part of the figure). On a Li substrate, particle world lines entangle, as expected of a superfluid, and even telling atoms apart is a difficult proposition; on the graphite substrate, on the other hand, after a relatively short time individual atoms begin to be clearly identifiable, as exchanges become suppressed and a regular (lattice) arrangement of atoms emerges. Concurrently, the superfluid fraction decreases (it is essentially zero, within statistical fluctuations, by the time the arrangement shown on the right side of Fig. \ref{compare} appears).

It is possible, and we have observed this in our simulations too, to have at times few isolated and resilient permutation cycles, which in a relatively small system can wind around the periodic
boundaries, giving rise to a finite superfluid signal. We attribute to this effect the observation of possible metastable superfluid phases in the grand canonical simulations of Ref. \cite{Happacher2013}. However, such a superfluid signal is spurious and disappears in the thermodynamic limit. A useful diagnostic tool is the calculated histogram of the frequency $P(n)$ of occurrence of exchange cycles including up to $n$ $^4$He atoms. 
In the superfluid phase, this is a smoothly  (exponentially) decaying function. On the other hand, if $P(n)=0$ for $n$ greater than a few 
(with the possible exception of isolated peaks at specific numbers, merely signaling the inability of the 
underlying sampling procedure to remove all permutation cycles), one is in the presence of a (crystalline) insulator. Our results clearly point to the latter scenario, i.e., they do not support the existence of a low-lying superfluid liquid-like phase, at variance with the contention of Ref. \cite {soliti2009}. We re-examine critically such a contention when discussing our ground state results in Section 3.2. 

\subsection{Ground state results} \label{sec:gsresults}
In the case of ground state simulations, we restricted ourselves to the case of commensurate coverage $\theta_0$.
The study of the ground state of the system described by Eq. \ref{u} was carried out using the following trial wave function:
\begin{equation}
    \Psi_T({\bf r}_1,{\bf r}_2,...{\bf r}_N) = \prod_{i<j}\ {\rm exp}  \biggl [-w(r_{ij}) \biggr ],
\end{equation}
where
\begin{equation}\label{twf}
    w(r)=\frac{\alpha}{1+\beta r^5}
\end{equation}
In the PIGS method \cite{Cuervo2005}, the true ground state $\Phi_0$ is approached by projecting it out of the initial trial wave function $\Psi_T$ as $\Phi_0$= lim$_{\Lambda\to\infty} \Phi(\Lambda) \equiv \lim_{\Lambda\to\infty}e^{-\Lambda\hat H}\Psi_T$. The values of the variational parameters in (\ref{twf}) are $\alpha=19$ and $\beta=0.12$ \AA$^{-5}$. 
 
 The most important aspect of this trial wave function is that it only includes pairwise correlations among helium atoms, accounting for the presence of a repulsive core at short interparticle distances in the interatomic potential $v$. Although it would be in principle possible to construct a trial wave function with additional terms reflecting the presence of a corrugated substrate, in a way that retains the intrinsic indistinguishability of helium atoms \cite{Boninsegni2001b}, the wave function utilized here has no built-in information concerning the presence of an attractive substrate. In other words,
 if one were to attempt a variational calculation the helium atoms would not remain close to the substrate, i.e., the whole system would simply evaporate. At the variational level, this is remedied by including, for example, a one-body term, confining helium atoms to the proximity of the substrate. In this work, we chose not to include any such term, relying instead on the projection algorithm to stabilize a $^4$He monolayer; we find that if $\Lambda\geq 0.125$ K$^{-1}$ a stable monolayer forms with no detectable atomic evaporation, and the $^4$He density profile in the direction perpendicular to the substrate is essentially indistinguishable from that computed at finite temperature, shown in Fig. \ref{static}.

 The reason for using a trial wave function of such a simple form is that it does not break translational invariance, and thus, by definition, it represents a fully superfluid state. 
 Since we aim to explore the possible presence of a finite superfluid signal in the ground state, we choose as the starting point of the projection a state that embodies the most favorable conditions for superfluidity,  searching for a residual superfluid signal in the ground state. As the true ground state breaks translational invariance due to the presence of the corrugated substrate,  it cannot be fully superfluid \cite{Leggett1970}.

 As stated in Section 2, we have observed numerical convergence of the estimates for all physical quantities, within statistical errors, for a value of the time step equal to that used in finite temperature simulations, namely $\tau=1.5\times 10^{-3}$ K$^{-1}$. For the ground state energy, on the other hand, we found that the value of the time step needed to observe convergence is as much as 10 times smaller. Our result for a projection time $\Lambda=1$ K$^{-1}$ is shown in Fig. \ref{energy}, and is consistent with the finite temperature estimate, within the statistical errors of the calculation (it is worth noting that the energy estimate yielded by PIGS for a finite projection time is a strict upper bound on the true ground state energy).
 
\begin{figure}[t]
\center
\includegraphics*[width=0.7 \linewidth]{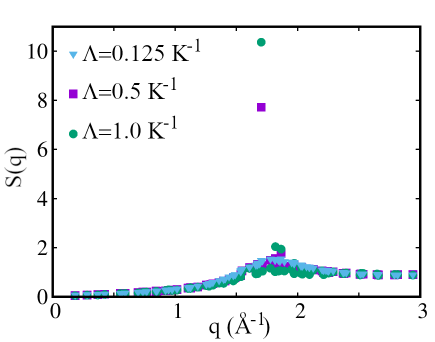}
\caption{Static structure factor computed by PIGS for three different projection times, for wave vectors in the plane of the substrate, showing the emergence of a sharp (Bragg) peak as the projection time is increased, signaling the onset of $\sqrt 3\times \sqrt 3$ crystalline order.}
\label{extrsq}
\end{figure}

 At this point, a general remark is in order. The ground state wave function of a Bose system, for the case of a Hamiltonian not breaking time-reversal symmetry, is real and positive. Consequently, a projection algorithm such as PIGS or DMC {\em necessarily} must converge to the true ground state, given a sufficiently long projection time, upon starting from a positive-definite trial wave function. One could imagine, in the case in which the ground state were, e.g., crystalline, that the projection time required might become {\em exceedingly} long if one started from a liquid-like trial wave function, and if there existed an excited state not possessing crystalline order, with energy very close to that of the ground state. One way to explore such an occurrence is by computing relevant quantities, indicative of the presence of order, as a function of projection time.
\\ \indent
 Fig. \ref{extrsq} shows the static structure factor $S(q)$ computed by PIGS for three different values of the projection time $\Lambda$. For a relatively short $\Lambda$, the projected state $\Phi(\Lambda)$ retains the physical properties of the trial wave function $\Psi_T$, including the lack of crystalline order. However, as $\Lambda$ increases a sharp peak in correspondence of the $\sqrt 3\times \sqrt 3$ crystalline order appears, with the height of the peak monotonically growing with $\Lambda$.\\ In other words, the projected state $\Phi(\Lambda)$ develops crystalline long-range order {\em irrespective of the fact that no such order is present in the trial wave function.} This is an all-important point, as it invalidates the contention made in Ref. \cite{soliti2009} that a low-lying liquid state very close in energy to the ground state exists, based on the presumption that a state projected out of a trial wave function not possessing crystalline order should retain the same physics, i.e., remain disordered. The result of Fig. \ref{extrsq} clearly shows that this is not the case; on the contrary, crystalline order is robust at low temperature, in this system, and no energetically competitive fluid state exists, a conclusion consistent with our findings at finite temperature.
\\ \indent
We now discuss the possible presence of a small but finite superfluid response in the ground state of the system at coverage $C_{1/3}$, to address the claim made in Ref. \cite{soliti2011} of a superfluid fraction $\rho_S=0.0067$ at zero temperature. 
No evidence of superfluidity has been seen in any finite temperature study, including this one \cite{Kwon2012,Happacher2013}, and even though the superfluid transition temperature can be expected to be relatively low, finite temperature simulations of small-size systems approaching that temperature failed to yield even a hint of developing superfluid order \cite{Happacher2013}.
\\ \indent
Within a ground state simulation method like PIGS, the superfluid fraction $\rho_s$ can be obtained from the long imaginary-time diffusion of particle world lines \cite{Zhang1995}, specifically as 
 \begin{equation} \label{sf}
   \rho_S={\rm lim}_{t \to\infty} D(t),\ \ {\rm with}\ \ D(t)=\frac{N}{2d\lambda}\frac{\expval{\qty[\vectorbold{R}_{\text{CM}}(t)-\vectorbold{R}_{\text{CM}}(0)]^2}}{t}
 \end{equation}
 where $\langle ... \rangle$ stands for average value, $d$ is the dimensionality (in this case $d=2$ as the system is essentially two-dimensional) and with $\vectorbold{R}_{\text{CM}}(t) = (1/N)\sum_{i=1}^{N}\vectorbold{r}_i(t)$, ${\bf r}_i(t)$ being the position of the $i$th particle at imaginary time $t$ along the world line, with $0 \le t \le \Lambda$. For a system that enjoys translational invariance, the estimator (\ref{sf}) trivially yields a value of 1, but if translational invariance is broken, as is the case in (\ref{u}), then $\rho_S < 1$ at $T=0$. \\ \indent Since the projection time $\Lambda$ is finite, the limit in (\ref{sf}) must be carried out by extrapolation, based on a fit of the diffusion curve $D(t)$ for a finite $\Lambda$ interval, in the $t\to\Lambda$ limit. We used here the formula proposed in Ref. \cite{Zhang1995}, i.e.,
 \begin{equation} \label{fit}
    D(t) = \rho_S + \frac{a}{t}\qty(1-\exp(-b{t}))
 \end{equation}
 where $a, b$ and $\rho_S$ are fitting parameters, $\rho_S$ representing the extrapolated superfluid fraction at infinite projection time. For a sufficiently large projection time $\Lambda$, one expects the values of the fitting parameters not to change, within their statistical uncertanties.
 
\begin{figure}[t]
     \centering
     \includegraphics[width=0.9\linewidth]{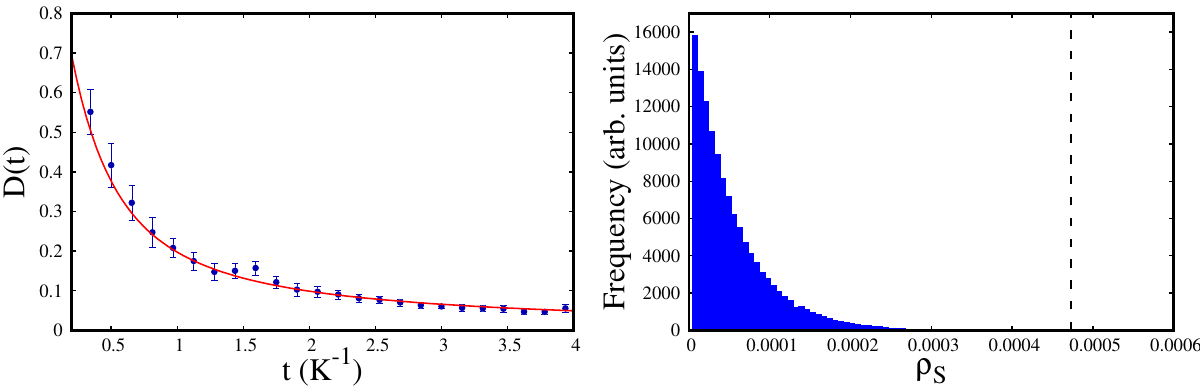}
     \caption{\textit{Left:} Fit to the diffusion curve at projection time $\Lambda = 4$ K$^{-1}$ using Eq. \ref{fit}. \textit{Right:} Computed probability distribution of values of $\rho_S$ emerging from the Metropolis random walk in parameter space utilized to fit the curve during the fitting procedure. The dotted line indicates the 99.99\% confidence limit.}
     \label{beta4}
 \end{figure}

We carried out such a fitting procedure by means of a Metropolis random walk through parameter space using the value of $\chi^2$ as the ``energy'' function; this approach offers access to the probability distributions of the values of the fitting parameters and allows for a reliable estimation of their uncertainties. 
Fig. \ref{beta4} illustrates the result of this procedure for $\rho_S$ in correspondence of a projection time $\Lambda=4$ K$^{-1}$.

Although the expression (\ref{fit}) aims at providing a good fit to the data only in the $t\to\Lambda$ limit, it turns out to be possible to obtain an acceptable fit, essentially over the whole imaginary time range, as shown by the left side of Fig. \ref{beta4}. The probability distribution for the value of the parameter $\rho_S$ of the fit, shown on the left side of Fig. \ref {beta4} places our estimate of the superfluid fraction in the ground state at less than $\sim 0.005$ with 99.99\% confidence, which is already well over an order of magnitude lower than the estimate of Ref. \cite{soliti2011}. We repeated this analysis for different values of $\Lambda$ but we never obtained a lower bound for $\rho_S$ greater than zero, with an acceptable statistical confidence. Thus, we conclude that our results are consistent with a null value of $\rho_S$ in the ground state, in accord with all finite temperature studies. Because this conclusion is established on a finite system, {\em a fortiori} it must hold the thermodynamic limit.
\\ \indent Once again, we wish to point out that by choosing a trial wave function that is translationally invariant we started out from the most favorable conditions for superfluidity. The fact that, given a sufficiently long projection time, the superfluid signal becomes essentially not measurable constitutes in our view strong evidence against the presence of superfluidity in the ground state. We attribute the disagreement between this conclusion and that of  Ref. \cite{soliti2011} to likely remnant variational bias in the DMC projection, quite likely arising from the finite population size, a problem already extensively documented in the literature \cite{Boninsegni2012b}.

 \section{Conclusion}
 We investigated the physics of a $^4$He monolayer adsorbed onto graphene substrate via Quantum Monte Carlo simulations both at zero and finite temperature. The goals of this study were to clarify the possible presence of (metastable) superfluid phases at commensurate and below commensurate ($C_{1/3}$) coverage and to clarify outstanding discrepancies between results and predictions yielded by different numerical studies. 
 \\ \indent
 The main conclusions of our study are that there are no (metastable) superfluid phases of the system at low temperature, either at $C_{1/3}$ coverage or below. In particular, our study lends no support to the contention, put forward in Ref. \cite{Kwon2012}, that the superfluid properties of the system below commensurate coverage are significantly affected by the choice of potential describing the interaction between a helium and a carbon atom (specifically, an isotropic potential stabilizing a superfluid phase below commensurate coverage). We find that the physics of the system is essentially the same for both types of potentials, and mimics that on a graphite substrate, i.e., no vacancy-induced supersolidity \cite {Dang2008} at low doping, with the formation of solid clusters at coverages below $C_{1/3}$.
 \\ \indent
 Our study also confirms the absence of any superfluid signal at commensuration, in agreement with all similar studies and with general arguments to the effect that no superfluid signal is seen at commensuration \cite{Dang2010}. Indeed, it is now accepted that only in the presence of multiple occupation of the unit cell (i.e., in cluster crystal) can supersolidity occurs at commensuration \cite{Boninsegni2012c}. We attribute the finite superfluid response reported at commensuration in Ref. \cite{soliti2011} to the (now understood) limitations of the technology utilized therein. 
 
 This work was supported by the Natural Sciences and Engineering Research Council of Canada. The computer codes utilized to obtain the results can be obtained by contacting the authors. The authors declare no conflict of interest.
\bibliographystyle{unsrt}
\bibliography{refs}
\end{CJK*}
\end{document}